\documentclass[mathleft
]{an}
\def\aap{A\&A}%
\let\aaps=\aap
\usepackage{natbib}
\bibpunct{(}{)}{;}{a}{}{,}
\usepackage{amssymb}
\usepackage{graphicx}
\usepackage{times}
\newcommand{\unit}[1]{\ensuremath{\:\mathrm{#1}}}
\newcommand{\muHz}{\unit{\mu Hz}}

\begin{document}

\Pagespan{000}{}
\Yearpublication{yyyy}%
\Yearsubmission{2010}%
\Month{mm}%
\Volume{VVV}%
\Issue{ii}%
\DOI{This.is/not.aDOI}%

\title{Current methods for analyzing light curves of solar-like stars}

\author{J. Ballot\thanks{\email{jballot@ast.obs-mip.fr}}}

\institute{Laboratoire d'Astrophysique de Toulouse-Tarbes, Universit\'e de Toulouse, CNRS, 14 av. E. Belin, 31400 Toulouse, France}

\received{1 May 2010}
\accepted{30 June 2010}
\publonline{later}


\abstract{CoRoT has allowed a quantitative leap for the solar-like-star seismology thanks to 5-month-long uninterrupted timeseries of high-precision photometric data. Kepler is also starting to deliver similar data. Now, several F and G main-sequence stars have been analyzed. The techniques developed to interpret light curves directly inherit from the experience got on the Sun with helioseismology. I describe in this review the methods currently used to analyze these light curves. First, these data provide an accurate determination of the stellar rotation rate. This is possible thanks to the magnetic activity of stars. The power spectra of light curves put also constraints on the stellar granulation, which can be directly compared to 3-D stellar atmosphere models; this shows still unexplained discrepancies. I then detailed a standard method for extracting p-mode characteristics (frequency, amplitude and lifetime). CoRoT has revealed unexpected short life times for F stars. Last, I also discuss errors and biases of mode frequencies, especially the ones due to the simplified description of the rotation generally used.}

\maketitle

\section{Introduction}
In solar-like stars, acoustic (p) modes can be excited whereas they are normally stable in regards to classical processes such as the well-known $\kappa$ mechanism. Damped modes are stochastically excited by the turbulent motions in their convective envelop. The resulting amplitudes are low. 
Excluding the Sun,
the first claim of a detection of solar-like p-mode oscillations has concerned Procyon. A power excess has been reported by \citet{Brown91}, lately confirmed by \citet{Martic99}. Nevertheless, the asteroseismology of solar-like stars has really started after the observations of $\alpha$~Cen~A by \citet{Bouchy01,Bouchy02} which has revealed a very clear comb structure in the oscillation spectrum. About 20 low-degree p modes in the spectrum have been identified.  After the Sun, it was the first opportunity to do seismological studies and to test models for this kind of stars
\citep[see for instance][]{Thevenin02,Toul03,Eggenberger04,Miglio05,Yildiz07}.

Today, several F-, G-, K-type dwarf and sub-giant stars have revealed p-mode oscillations thanks to accurate radial-velocity measurements performed with high-resolution spectrograph. To avoid the day-night alternation which generates gaps in data and then windowing effects, one can run multi-site campaigns, such as those dedicated to Procyon \citep{Arentoft08,Bedding10}, or use space-based instruments which can follow continuously targets during several weeks, several months, or even more.
P modes in solar-like stars have then been accurately measured by CoRoT \citep{Michel08} and Kepler \citep{Chaplin10}.
Contrary to ground-based observations measuring radial velocities, space-based missions perform high-precision photometry measurements. Acoustic modes generate small fluctuations in the light curve which are analyzed.

Thanks to their long duration, their high duty cycle, and their good signal-to-noise ratio, data sets obtained by space missions can be analyzed with techniques inherited from helioseismology. In this paper, we focus on the classical way to analyze such long photometry data sets for solar-like stars. We will illustrate these standard analysis techniques with recent results obtained with CoRoT.

\section{Non-seismic information in seismic data}

High-precision photometry timeseries reveal more than acoustic modes in solar-type stars. Indeed, the luminosity fluctuates not only because of intrinsic oscillations, but also because of magnetic activity and surface motions.
We discuss in this section what is the information we can obtained about the rotation of stars and their surface convection.

\subsection{Rotation}

The first visible expression of the magnetic activity of the Sun is the presence of spots visible on the photosphere. Spots are good trackers -- and historically the first ones -- of the solar rotation. If we use this solar description as a paradigm, such spots should be present on the surface of solar-like stars with convective envelop which develop a magnetism similar to the Sun. The presence of spots, but also of bright areas called plages, creates modulations in the light curve with a period equal to the rotation period. 
Indeed, a magnetic structure creates such a modulation if it is alternately visible or hidden on the far side. Due to the inclination $i$ of the stellar rotation axis, structures too close to the poles are always visible or hidden and do not produce a modulation. Moreover, the lifetime of structures must be long enough, compared to the rotation period, to create repetitive patterns. Such modulations can be found in the power spectrum of stellar light curves. Contrary to rotation velocities measured in spectroscopy with line broadening, these direct measurements of rotation periods is decoupled from $\sin i$.

\begin{figure}
\includegraphics[width=\linewidth]{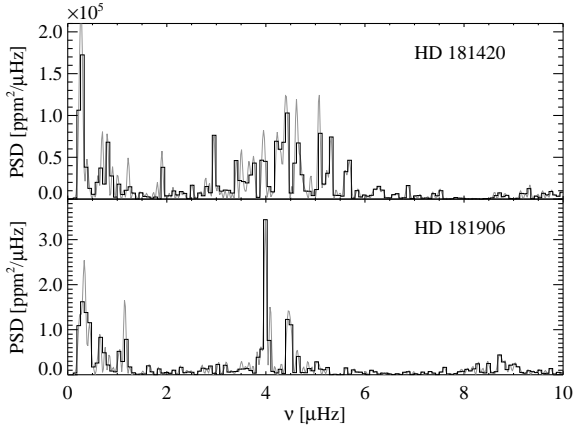}
\caption{Power spectrum in a low frequency range of the light curves of two CoRoT targets HD~181420 \citep[after][]{Barban09} and HD~181906 \citep[after][]{Garcia09} showing the rotation frequency. In gray, the spectrum is oversampled with a factor of 4.}
\label{fig:rotdif}
\end{figure}

Figure~\ref{fig:rotdif} shows the power spectrum in the low frequency range for two CoRoT targets. Clear peaks appear as the signature of the rotation, but patterns are different for the two stars. For HD~181420 \citep{Barban09}, the pattern is broad with numerous peaks, whereas for HD~181906 \citep{Garcia09}, the power is concentrated in two peaks.
We interpret the broadening as an effect of differential rotation in latitude. A spot tracks the rotation at the latitude it has appeared. When there are spots at various latitudes and a noticeable differential rotation, we observe a broad pattern in the power spectrum. On the other hand, a narrow peak can be due either to a weak differential rotation rate or to a small number of spots.
For deeper analyses, it is also possible to directly fit the light curve with spot models \citep[see for instance][]{Lanza06,Mosser09Spot}.

It is worth noting that these observations give an information on the presence of magnetic structures, even on small scales. It is then possible to have an indication of the stellar magnetic activity, according to the coverage of the surface with spots or plages. This information is complementary to observations of large-scale magnetic fields with spectropolarimetry \citep[e.g.][]{Petit08}.

\subsection{Power spectrum of convection and background}\label{ssec:bg}

\begin{figure}
\includegraphics[width=\linewidth]{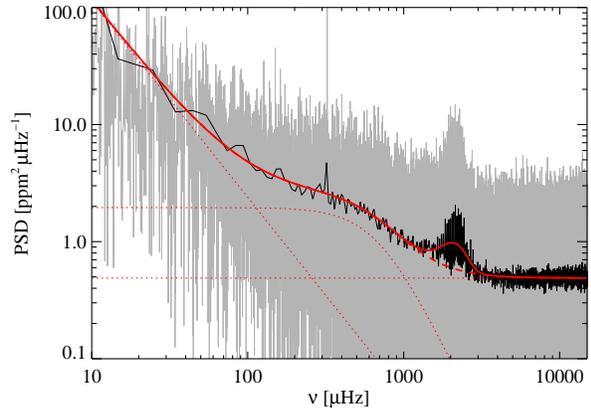}
\caption{Power spectrum of a CoRoT target in gray. In black, the same plot smoothed over 100 points. Red dotted lines show the three components of the background (see text) fitted on the data, the red dashed line is the sum. The solid red line includes the p-mode contribution modeled with a Gaussian function.}
\label{fig:bg}
\end{figure}

When we analyze the power spectrum of solar-like stars at higher frequency, we identify several components illustrated in Fig.~\ref{fig:bg}.
The p-mode contribution is visible in this example as a bump around 2000\muHz. The rest forms the background we decompose classically in three components as following:
\begin{enumerate}
\item a white noise, mainly due to photon noise, dominating the spectrum at the highest frequencies;
\item a low-frequency component, which can be modelled with a power law, resulting to activity effects and slowly varying trends;
\item the contribution of the surface convection, i.e. the granulation, described with a model of \citet{Harvey85}:
\begin{equation}
B_c(\nu)=\frac{\tau_c\sigma_c^2}{1+(2\pi\tau_c\nu)^{\alpha_c}},
\end{equation}
where $\nu$ is the frequency, and $\sigma_c$, $\tau_c$ and $\alpha_c$ are 3 free parameters. $\tau_c$ represents the characteristic time of the granulation.
\end{enumerate}
Some authors \citep[see][]{Karoff10} add also another contribution due to faculae. Faculae are brights points visible at the solar surface that should also exist on other solar-type stars.

For more than 10 years, the granulation patterns observed at the surface of the Sun are correctly reproduced by 3-D models of the solar atmosphere performed with radiative hydrodynamics code \citep{SteinNordlund98}. Similar models have been performed for other stars. For instance, \citet{Ludwig09} have performed models for the CoRoT target HD 49933 and they have compared the power spectrum deduced from their simulations to the CoRoT observations. When the granulation timescale is correctly reproduced, the amplitude of the convective background is significantly different within a factor of 2--3 for this star. This discrepancy has still to be explained.
All this new bench of data offers an interesting opportunity to constrain this kind of 3-D models, which have already proved their values with their capability to consistently reproduce solar absorption line profiles \citep[e.g.][and following works]{Asplund00}.

\section{Mean seismic information}

For faint stars, due to the low signal-to-noise ratio, p modes are generally seen as a small power excess over the background, but not as identifiable peaks. Nevertheless, it is still possible to extract seismic information.
The most obvious information are $\nu_{\mathrm{max}}$, $A_{\mathrm{max}}$, and  $\Delta\nu$, which denote respectively the frequencies and amplitudes of the most powerful modes, located at the center of the power excess bump, and the large frequency separation \citep[see for instance the CoRoT target HD~175726 and HD~170987 in][]{Mosser09_HD175726,Mathur10}.
However, the estimates of $A_{\mathrm{max}}$ are to be cautiously considered. Due to a possible overestimation of the background contribution, \citet{Mathur10pipe} have shown that, using their standard method, $A_{\mathrm{max}}$ is systematically underestimated.
Detecting a power excess is not sufficient: it has to be shown that it is produced by the p modes and does not originate from another phenomena such as faculae, as mentioned in Sect.~\ref{ssec:bg}. This problem is discussed in \citet{Mathur10}. To have a solid clue of its seismological origin, we must detect a periodic spacing in this power excess. Such a period spacing proves the presence of a comb structure hidden in the noise. By performing the autocorrelation --or the spectrum-- of the power spectrum restricted in a range around $\nu_{\mathrm{max}}$, we then determine the mean large separation $\langle\Delta\nu\rangle$.

In the past few years, several pipelines have been developed to extract these global seismic parameters in an automatic way \citep{Huber09,Hekker10,Kallinger10,Mathur10pipe}. Automatic procedures are needed today to efficiently exploit the observations of big sets of stars as those produced by Kepler.

The determination of the mean large separation $\langle\Delta\nu\rangle$, coupled with effective temperature measurements, provides an estimate of the stellar radius within an accuracy of a few percents as discussed in \citet{Stello09}. Of course, we then assume that stellar models are perfectly known. Such constraints do not improve our view of stellar physics nor refine our stellar evolution models, but are really useful for instance to characterize planet transits.

To go beyond the mean large spacing $\langle\Delta\nu\rangle$, the autocorrelation function (ACF) with a narrow filter is a valuable tool. The ACF can be computed as the Fourier spectrum of the filtered oscillation spectrum \citep{Roxburgh06}. If the bandpass filter applied to the oscillation spectrum is narrow enough, by scanning the p-mode frequency range, it is possible to recover variations of $\Delta\nu$ with frequency \citep[for more details on this technique, see][]{Roxburgh09,Mosser09ACF}. If the noise is sufficiently low and the frequency resolution sufficiently fine, the autocorrelation function recovers the oscillation in $\Delta\nu(\nu)$ produced by the second helium ionisation \citep[e.g.][]{Monteiro05}. Such an information provides a valuable quantity to constrain models more precisely.

\section{Individual p-mode determination}

\subsection{Modeling the spectrum}
In the best cases, when the signal-to-noise ratio is sufficiently high, p modes are individually visible in the spectrum. Acoustic modes in solar-like stars are considered as damped modes continuously and stochastically re-excited by the near-surface convection \citep[e.g.][]{Kumar88}. The power spectrum of such a mode is a Lorentzian profile multiplied by a stochastic noise following a exponential distribution. This approach is valid for the Sun. Although asymmetric Lorentzian profiles are generally considered in the solar case \citep{Abrams96}, nowadays, we totally neglect profile asymmetries for other stars due to the coarser frequency resolution and the lower signal-to-noise ratio. However, with longer observations, such as the Kepler long runs, the asymmetry will become a relevant parameter to fit, and will be needed to avoid biases in frequency determination. We also assume a low rotation rate $\Omega$, without noticeable differential rotation. The rotation is treated as a first-order perturbation, and the Coriolis force effect is neglected since the mode frequencies are large compared to the rotation frequency \citep{Ledoux51}. This assumption is discussed in Sect.~\ref{sec:rot}. Thus, the frequency of a mode with radial order $n$, degree $l$ and azimuthal order $m$ is $\nu_{n,l,m}=\nu_{n,l}-m\nu_s$, where $\nu_{n,l}$ is the frequency in absence of rotation and $\nu_s\approx\nu_{\mathrm{rot}}\equiv\Omega/2\pi$ is the rotational splitting.

Thus, we model the power spectrum $S(\nu)$ of a light curve as following
\begin{equation}
S(\nu) = \left(B(\nu) + P(\nu) \right)\times {\cal N}(\nu)
\end{equation}
where $B$ is the background described in Sect~\ref{ssec:bg}, ${\cal N}$ a random function following a exponential law, and 
\begin{equation}
P(\nu) = \sum_{n,l}\sum_{m=-l}^l \frac{H_{n,l} a_{l,m}(i)}{1+[2(\nu-\nu_{n,l}+m\nu_s)/\Gamma_{n,l}]^2}.\label{eq:spectrum_model}
\end{equation}
$H_{n,l}$ are the modes heights, $\Gamma_{n,l}$ their widths -- linked to the lifetimes $\tau_{n,l}=(\pi\Gamma_{n,l})^{-1}$ -- and $a_{l,m}(i)$ is the amplitude ratio, expressed as a function of the inclination angle:
\begin{equation}
a_{l, m}(i)=\frac{|l-m|!}{|l+m|!}(P_l^m(\cos i))^2
\end{equation}
where $P_l^m$ are the associated Legendre functions \citep[e.g.][]{Gizon03}. Moreover, we generally assume that the width is a slow function of the only frequency $\Gamma_{n,l}=\Gamma(\nu_{n,l})$. Similarly, we assume for the height, 
\begin{equation}
H_{n,l}=(V_l/V_0)^2H_0(\nu_{n,l}), 
\end{equation}
where $V_l$ is the mode visibility \citep[for a generally discussion, see][]{Toutain93}. In other words, the energy equipartition between modes of close frequency is assumed. $V_l$ is rapidly decreasing with $l$ and becomes very small for $l>2$. Thus, in practice, only modes $l=0,1$ and 2  (and, sometimes, $l=3$) are considered. Under very general assumption, it is worth noting we have the relation
\begin{equation}
\sum_k V^2_{l=2k}=\sum_k V^2_{l=2k+1}\label{eq:sumvis}
\end{equation}
A demonstration is proposed in appendix.

To recover the values of the free parameters, the model is fitted to the data using a maximum likelihood estimator (MLE), as it has been done for the Sun \citep[see method in][]{Appourchaux98}. Nevertheless, it is now very common to consider Bayesian priors on the parameters and then to use maximum a posteriori estimators \citep[MAP, e.g.][]{Gaulme09}. In Bayesian approaches, Monte Carlo Markov chains (MCMC) have shown their usefulness on CoRoT data \citep{Benomar09}: when classical estimators only return the value of a maximum of probability (which can be only local in the large parameter space to explore), MCMC can return the marginal probability density function (PDF) for each parameter. PDFs sometimes appear to be non-normal, or even multimodal.

\subsection{Errors on parameter estimation}
The best way to determine the errors of parameters is to derive them from the PDF computed with MCMC. When one uses a MLE, the errors are estimated from the Hessian matrix \citep{Appourchaux98}. These estimations of errors are based on the Kramer-Rao theorem. Thus we have to keep in mind the following important points:
\begin{enumerate}
\item it is only a lower limit of the error;
\item it is only asymptotically valid, so the PDF is supposed to be normal \citep[that is not always the case with several months of data, especially for $i$ and $\nu_s$, see][]{Ballot08};
\item the Hessian matrix should be computed for the real values of the parameters, and not for the estimated values as done during a real fit.
\end{enumerate}

Some general properties of errors have been studied last years. Concerning the inclination and the splitting, 
\citet{Ballot06} have shown the strong correlation between the determination of $i$ and $\nu_s$, when $\nu_s\lesssim\Gamma$ (this is generally the case). Thus the errors of $i$ and $\nu_s$ become huge, but projected splittings $\nu_s^*=\nu_s\sin i$ are better determined.

The Kramer-Rao limit for $\nu_{n,l}$ has be determined first for a single Lorentzian profile by \citet{Libbrecht92}. It has been generalized for multiplets in the solar configuration ($i=90\degr$) by \citet{Toutain94} and for any inclination by \citet{Ballot08}. The errors of $\nu_{n,l}$ are:
\begin{equation}
\sigma_{\nu_{n,l}}=\sqrt{\frac{1}{4\pi}\frac{\Gamma_{n,l}}{T}f_l(\beta_{n,l},x_s,i)},\label{eq:errnu}
\end{equation}
with $T$ the observation duration, $\beta_{n,l}=B(\nu_{n,l})/H_{n,l}$ the noise-to-signal ratio, and $x_s=2\nu_s/\Gamma_{n,l}$ the reduced splitting. The functions $f_l$ are described in \citet{Ballot08} and varies typically between 0.5 and 1.5 for CoRoT standard targets.

\subsection{The F-type star HD~49933}
The first solar-like star observed by CoRoT was HD~49933 \citep{Appourchaux08}. It has revealed modes broader than expected. Mode widths are larger than the small separation 0--2, thus $l=2$ modes are blended with $l=0$ modes. Moreover, due to Eq.~\ref{eq:sumvis}, the power of $l=1$ modes is very close to the sum of the power of nearby $l=0$ and $l=2$ modes. The modes are then not easily identified. With 60 days of data, the mode identification favored by \citet{Appourchaux08} has appeared to be wrong. It has been possible to undoubtedly distinguish even and odd modes with longer observations \citep{Benomar09}.

\begin{figure}
\includegraphics[width=\linewidth]{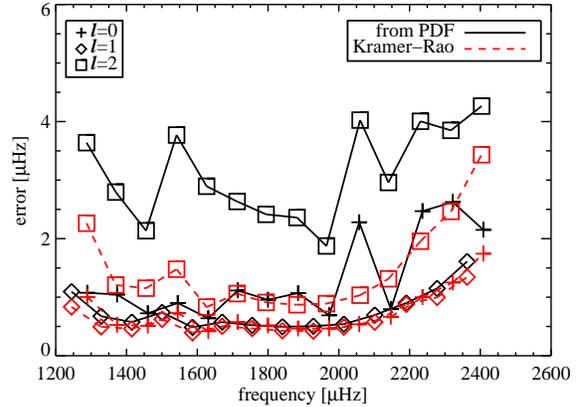}
\caption{Errors of mode frequencies $\nu_{n,l}$ for HD~49933 after 180 days of observations. Black lines/symbols correspond to the real errors determined on the PDF by \citet{Benomar09}. Red lines/symbols show Kramer-Rao limits determined with Eq.~\ref{eq:errnu}.}
\label{fig:errnu}
\end{figure}
Such high mode widths has consequences on frequency accuracies, according to Eq.~\ref{eq:errnu}. Moreover, errors are increased by the blending of $l=0$ and 2 modes. Indeed, Eq.~\ref{eq:errnu} is derived for isolated modes; when two modes are blended, it appears correlations in the determination of the two mode frequencies, that tends to increase the errors. This is illustrated on Fig.~\ref{fig:errnu}. For HD~49933, the real errors of $l=1$ mode frequencies, determined from the PDF, are almost equal to the Kramer-Rao limit given by Eq.~\ref{eq:errnu} ($\sim 0.5\muHz$). However, the errors for even modes, especially for $l=2$ modes, are increased.
Accurate determinations of $l=2$ modes for this kind of F stars appear to be very difficult, and require year-long observations, as Kepler will do.

\subsection{Mode lifetimes}

Mode lifetimes appear to decrease when effective temperatures $T_{\mathrm{eff}}$ increase. Using theoretical models for damping processes and some seismic observations based on radial velocity measurements, \citet{Chaplin09} have found a strong dependence $\Gamma\propto T_{\mathrm{eff}}^4$. Recently, by considering CoRoT observations, \citet{Baudin10} have even found a stronger dependence $\Gamma\propto T_{\mathrm{eff}}^{14}$. This is different from what they have found for solar-like oscillations in red giants where $\Gamma$ is weakly sensitive to $T_{\mathrm{eff}}$.

Such a strong dependence on $T_{\mathrm{eff}}$ has still to be explained. Progresses should be done to understand the damping processes. Nowadays, there is not clear consensus and different mechanisms have been proposed.
\citet{Goldreich91} have shown that
radiative losses and turbulent viscosity play a major role in
mode damping, whereas \citet{Gough80} or \citet{Balmforth92}
found that the damping is dominated by the modulation of turbulent
pressure. More recently, \citet{Dupret06} suggested the dominant contribution is
the perturbation of the convective heat flux.

\section{Effects of rotation on the oscillations}\label{sec:rot}
Perturbative treatments are valid only for low rotation rates. Is it the case in practice? As an illustration, we consider HD~49933. Its rotation frequency is $\nu_{\mathrm{rot}}\approx 3.4\muHz$, i.e. $\Omega\approx 0.05\Omega_K$ with $\Omega_K=\sqrt{GM/R^3}$ the Keplerian frequency, derived from estimated mass $M$ and radius $R$ of the star. According to \citet{Reese06}, in this regime, perturbative treatments of the rotation are still valid, but second order terms must be considered. Frequencies are then expressed
\begin{equation}
\nu_{n,l,m}=\nu_{n,l}+(1-C_{n,l})m\nu_{\mathrm{rot}} + (D^1_{n,l}+m^2D^2_{n,l})\nu_{\mathrm{rot}}^2\label{eq:saio}
\end{equation}
according to \citet{Saio81}. I have crudely estimated perturbative coefficients $C_{n,l}$, $D^1_{n,l}$ and $D^2_{n,l}$ from a polytropic model by using results from \citet{Reese06} (see Table~\ref{tab:coef}). 
As expected, the Coriolis correction $C_{n,l}$ is very small and can be neglected. However the second-order terms modify the frequencies by several~\muHz. Multiplets are shifted, and they are no more symmetric relative to the $m=0$ component. Moreover, this estimation is optimistic, since the values used here correspond to low-order modes ($n\approx 10$): for higher orders, second-order terms should even be greater.

\begin{table}
\caption{Perturbative coefficients representative of a HD~49933-like star used in this paper. $C$ is a natural number, $D^1$ and $D^2$ are expressed in~\unit{\mu Hz^{-1}}. These coefficients have been estimated by using results from \citet{Reese06} and correspond to $n\approx 10$.}\label{tab:coef}
\centering
\begin{tabular}{lccc}
\hline
\hline
$l$ & $C$     & $D^1$   & $D^2$  \\
\hline
0   & --      & $-0.15$ &  --    \\
1   & $-0.01$ & $-0.10$ & $-0.10$\\
2   & $-0.01$ & $-0.13$ & $-0.02$\\
\hline
\end{tabular}
\end{table}

\begin{figure}
\includegraphics[width=\linewidth]{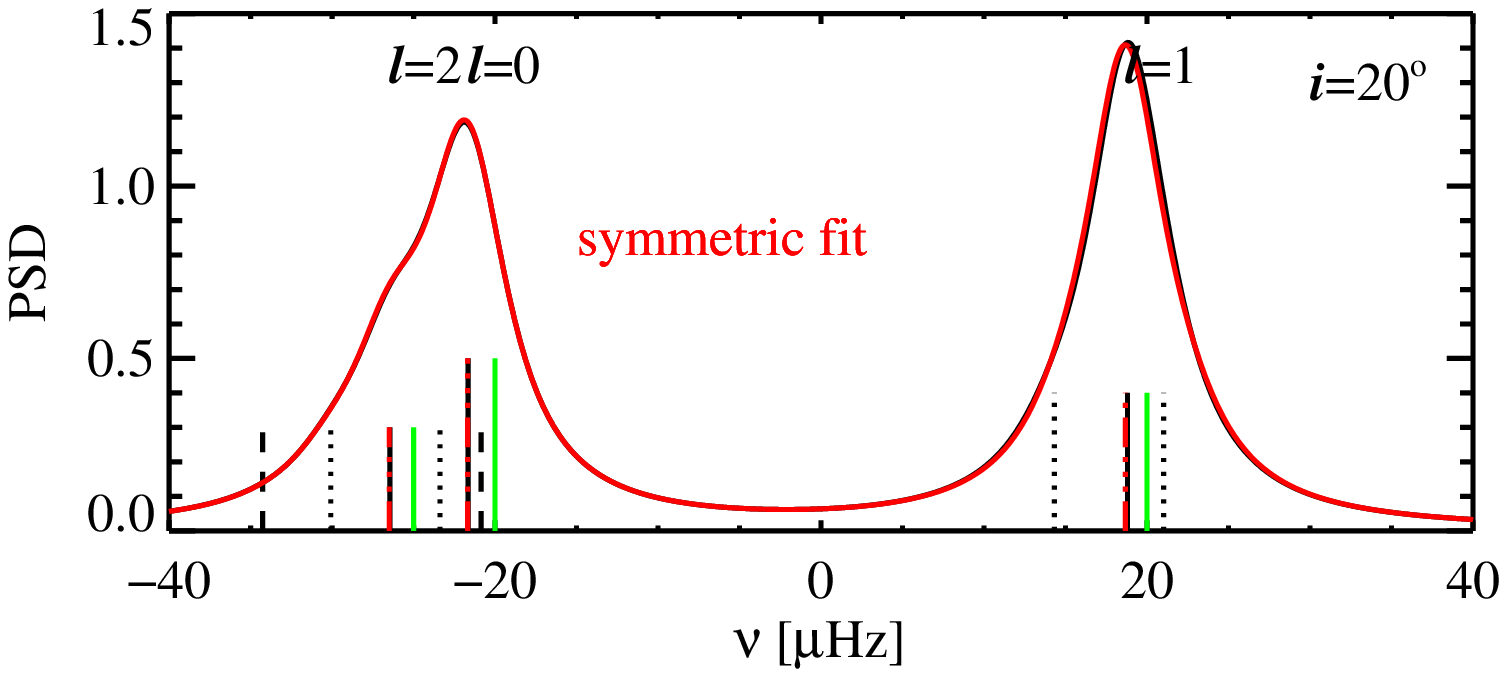}
\includegraphics[width=\linewidth]{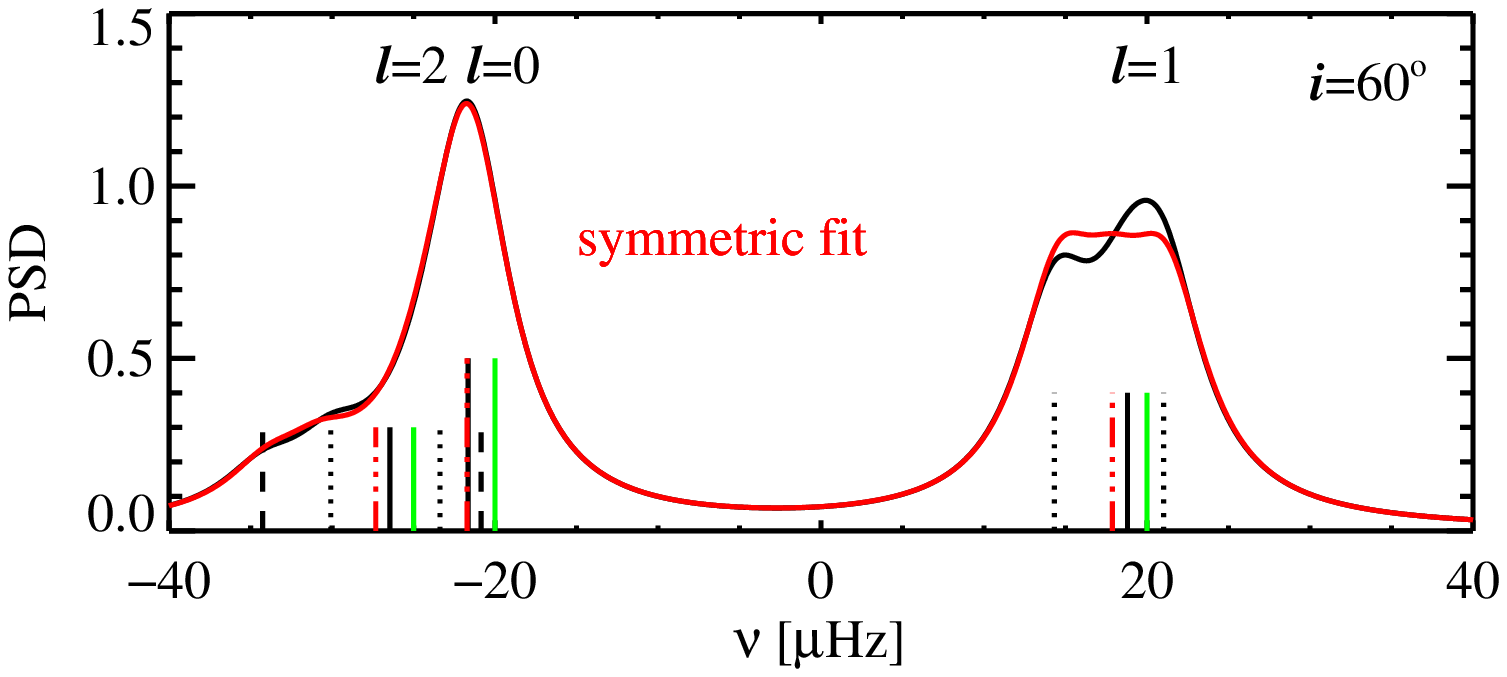}
\includegraphics[width=\linewidth]{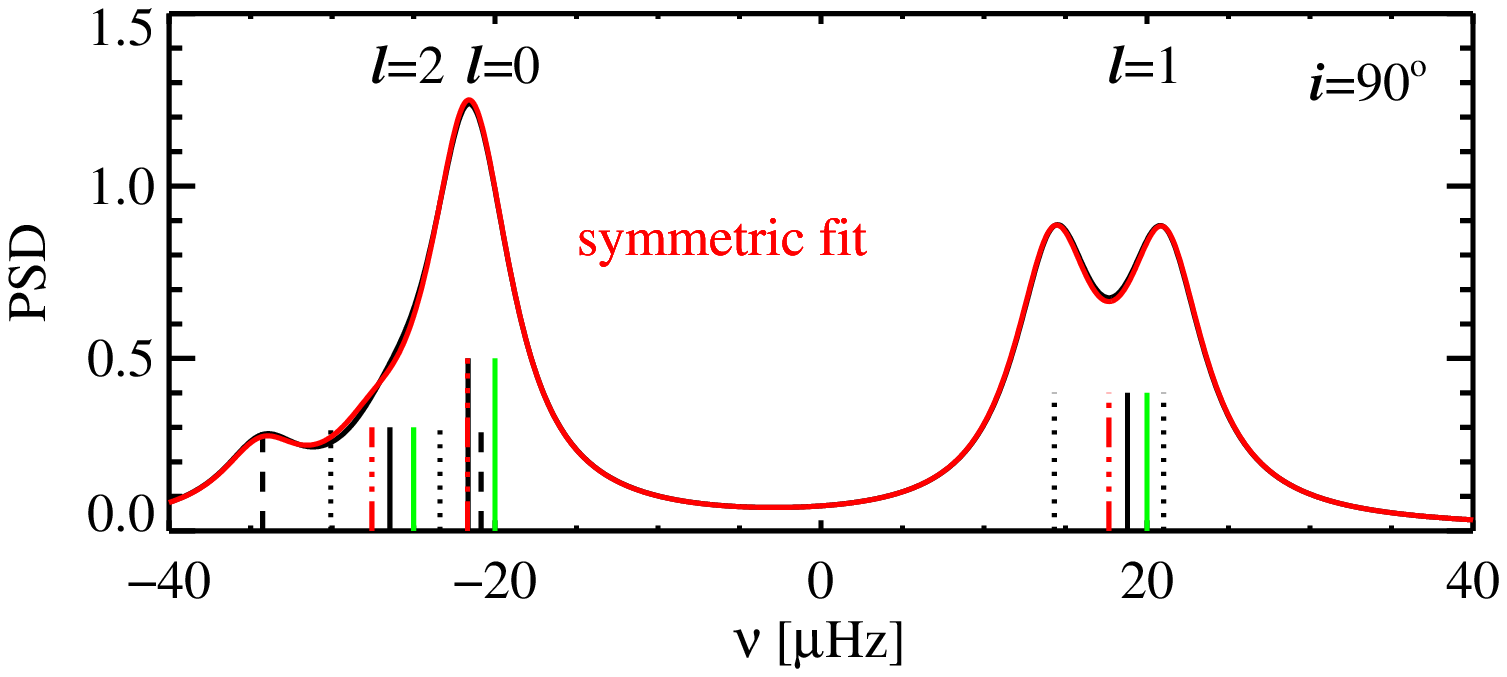}
\caption{Synthetic spectra (black lines) showing three consecutive modes $l=2,0$ and 1, for HD~49933-like stars with different inclination angles $i=20$, 60 \& 90\degr (top to bottom panels). Mode frequencies are computed with Eq.~\ref{eq:saio}. Green ticks indicate the frequencies without rotation, black ticks the frequencies with rotation (solid, dotted and dashed ticks correspond to $|m|=0,1$ \& 2 components). Red lines show the fit of Eq.~\ref{eq:spectrum_model} and red ticks indicate the fitted frequencies.}
\label{fig:effect_rot}
\end{figure}

Since this asymmetry is not taken into account in the fitted spectrum model (Eq.~\ref{eq:spectrum_model}), the fitted parameters are biased. To estimate these biases, I have computed synthetic spectra using the 2nd-order perturbative formulation (Eq.~\ref{eq:saio}) and performed classical fits. Results are shown in Fig.~\ref{fig:effect_rot}.

Biases for frequencies $\nu_{n,l}$ vary according to the inclination, typically from $\sim 1.3$ to 2.6\muHz. For low angles ($i\lesssim 30\degr$), fits are dominated by the $m=0$ components, so the fitted frequencies are $\nu_{n,l}^{(fit)}\approx \nu_{n,l}+D^1_{n,l}\nu_{\mathrm{rot}}^2$. For high angles ($i\gtrsim 80\degr$), spectra are dominated by the sectoral modes $|m|=l$, thus the fitted frequencies are $\nu_{n,l}^{(fit)}\approx \nu_{n,l}+(D^1_{n,l}+l^2D^2_{n,l})\nu_{\mathrm{rot}}^2$.
Fitted rotational splittings $\nu_s$ deviate also from the value of $\nu_{\mathrm{rot}}$, especially at low angles: the bias is around 0.3\muHz\ for our case with $i=20\degr$. However, for $i\gtrsim 80\degr$, the bias vanishes: the sectoral modes dominate and they are still separated by $2l\nu_{\mathrm{rot}}$.
Concerning the other parameters ($i$, $H$ and $\Gamma$), the biases are small and negligible relatively to the errors.

Biases are significant for frequencies and splittings in stars like  HD~49933. Second-order terms must then be considered. Ideally, it should be included in the fitted models, but it increases the number of free parameters and risk destabilising the fit. Nevertheless it must be taken into account for the interpretation of frequencies.

\section{Conclusion}

Last years, seismic data of unparalleled quality have been produced for solar-like stars, thanks to CoRoT, and now, thanks to Kepler.
The delivered high-precision photometric timeseries put a lot of news constraints on stars.
First, we get accurate measurements of rotation rates and good indications for magnetic activities. Next, the background of the spectrum as well as mode amplitudes and lifetimes constrain the turbulent dynamics of the near-surface layers, and it offers new capabilities to test 3-D stellar atmosphere models. The extracted frequencies must now be interpreted in terms of constraints on the stellar structure, especially to improve the treatment of mixing processes. Nevertheless, the rotation has to be carefully considered, not only because of the effects induced on the structure and the evolution of stars, but also because frequencies are significantly affected even for typical F main-sequence stars.

\acknowledgements
The author acknowledges the support of the Agence National de la Recherche through the SIROCO project. This work was supported by the European
Helio- and Asteroseismology Network (HELAS), a major international
collaboration funded by the European Commission's Sixth
Framework Programme.

\appendix
\section{Odd and even mode visibilities}\label{App}
The visibility $V_l$ is given by the relation \citep{Gizon03}:
\begin{equation}
  \label{eq:Vl}
  V_l=2 \pi \int_0^1 Y_l^0(\mu) W(\mu) \mu d\mu
\end{equation}
where $Y_l^m$ denotes the spherical harmonics and $W$ is a function indicating the contribution to the total light flux of an element of the stellar disk; we have assumed that $W$ depends only on the distance to the limb $\mu$.
We rewrite the equation
\begin{equation}
  V_l=\sqrt{\pi(2l+1)} \int_0^1 P_l(\mu) f(\mu) d\mu
\end{equation}
with $f(\mu)=W(\mu)\mu$ and $P_l$ the Legendre polynomials.
We define $f_e(\mu)$ the \emph{even} function over the interval $[-1,1]$ such that $f_e(\mu)=f(\mu)$ $\forall \mu \in [0,1]$ and $f_o(\mu)$ the \emph{odd} function over the interval $[-1,1]$ such that $f_o(\mu)=f(\mu)$ $\forall \mu \in [0,1]$.

Legendre polynomials form a orthogonal basis of the Hilbert space $\mathbf{L}_2([-1,1])$ where is defined a dot product 
\begin{equation}
\langle f, g\rangle=\int_{-1}^{1} f(x) g(x) dx.
\end{equation}
Legendre polynomials verify the relation
\begin{equation}
  \langle P_l, P_{l'}\rangle=\frac{2 \delta_{l,l'}}{2l+1}
\end{equation}
As $f_e$ and $f_o$ are obviously elements of $\mathbf{L}_2([-1,1])$, we decompose them on the Legendre basis and take into account their parity:
\begin{eqnarray}
  \label{eq:decomp}
  f_e(x)&=&\sum_{k=0}^{\infty} \alpha_{2k} P_{2k}(x)\\
  f_o(x)&=&\sum_{k=0}^{\infty} \beta_{2k+1} P_{2k+1}(x)
\end{eqnarray}
Thus for $l=2k$ and $l=2k+1$, we obtain respectively
\begin{eqnarray}
  V_{2k}&=&\sqrt{\pi(4k+1)}\, \frac{1}{2}\int_{-1}^1 P_{2k}(\mu) f_e(\mu) d\mu\\
        &=&
\sqrt{\frac{\pi}{4k+1}} \alpha_{2k}\label{eq:ve}\\
  V_{2k+1}&=&\sqrt{\pi(4k+3)}\, \frac{1}{2}\int_{-1}^1 P_{2k+1}(\mu) f_e(\mu) d\mu\\
        &=&
\sqrt{\frac{\pi}{4k+3}} \beta_{2k+1}\label{eq:vo}
\end{eqnarray}
We consider now $\langle f_e, f_e \rangle$ the square of the norm of $f_e$. On one hand, we get
by using its decomposition on the Legendre basis
\begin{equation}
\langle f_e, f_e \rangle= \sum_{k=0}^{\infty} \alpha_{2k}^2 \frac{2}{4k+1}.
\end{equation}
On the other hand,
\begin{equation}
\langle f_e, f_e \rangle= \int_{-1}^{1} f_e(x)^2 dx = 2 \int_{0}^{1} f(x)^2 dx
\end{equation}
then
\begin{equation}
  \sum_{k=0}^{\infty}  \frac{\alpha_{2k}^2}{4k+1} = \int_{0}^{1} f(x)^2 dx.\label{eq:a2}
\end{equation}
By considering $\langle f_o, f_o \rangle$, we show similarly that
\begin{equation}
  \sum_{k=0}^{\infty}  \frac{\beta_{2k+1}^2}{4k+3} = \int_{0}^{1} f(x)^2 dx.\label{eq:b2}
\end{equation}
Hence we have
\begin{equation}
  \sum_{k=0}^{\infty}  \frac{\alpha_{2k}^2}{4k+1} =
  \sum_{k=0}^{\infty}  \frac{\beta_{2k+1}^2}{4k+3} = \int_{0}^{1} f(x)^2 dx, \label{eq:a2b2}
\end{equation}
Finally, by injecting Eqs~\ref{eq:ve} \& \ref{eq:vo} in Eq.~\ref{eq:a2b2}, we get
\begin{equation}
  \sum_{k=0}^{\infty} V_{2k}^2 =
  \sum_{k=0}^{\infty} V_{2k+1}^2 = \pi \int_0^1W(\mu)^2\mu^2 d\mu.
\end{equation}

\end{document}